\documentclass[review]{elsarticle}

\usepackage{amsmath,amssymb,bm,bbm}
\usepackage{amsthm}
\usepackage{graphicx,psfrag,epsf}
\usepackage{enumerate}
\usepackage{float}
\usepackage{subcaption}
\usepackage{enumerate}
\usepackage{dsfont}
\usepackage{natbib}
\usepackage{url} 
\usepackage{geometry}
\usepackage{marginnote}
\usepackage{xcolor}
\usepackage{comment}
\usepackage[colorinlistoftodos]{todonotes}

\usepackage[utf8]{inputenc}
\newtheorem{theorem}{Theorem}[section]

\newcommand{\N}{\mathrm{N}}

\journal{Journal of \LaTeX\ Templates}

\biboptions{authoryear}

\begin{document}

\begin{frontmatter}

\title{Bayesian Analysis of Nonparanormal Graphical Models Using Rank-Likelihood}

\author{Jami J. Mulgrave \fnref{myfootnote}}
\author{Subhashis Ghosal}
\address{Department of Statistics\\
       North Carolina State University\\
       Raleigh, NC 27695-8203, USA}
\fntext[myfootnote]{The authors gratefully acknowledge the National Science Foundation (NSF) Graduate Research Fellowship Program Grant No. DGE-1252376, the National Institutes of Health (NIH) training grant GM081057, NSF grant DMS-1732842, and NSF grant DMS-1510238.}




\begin{abstract}
Gaussian graphical models, where it is assumed that the variables of interest jointly follow a multivariate normal distribution with a sparse precision matrix, have been used to study intrinsic dependence among variables, but the normality assumption may be restrictive in many settings. A nonparanormal graphical model is a semiparametric generalization of a Gaussian graphical model for continuous variables where it is assumed that the variables follow a Gaussian graphical model only after some unknown smooth monotone transformation. We consider a Bayesian approach for the nonparanormal graphical model using a rank-likelihood which remains invariant under monotone transformations, thereby avoiding the need to put a prior on the transformation functions.  On the underlying precision matrix of the transformed variables, we consider  a horseshoe prior on its Cholesky decomposition and use an efficient posterior Gibbs sampling scheme. We present a posterior consistency result for the precision matrix based on the rank-based likelihood.  We study the numerical performance of the proposed method through a simulation study and apply it on a real dataset.
\end{abstract}


\end{frontmatter}


\section{Introduction}
\label{sec:intro}
Graphical models are useful mathematical tools that model complex dependence relationships between variables.  Under the Gaussianity assumption, the graph of relations is completely determined by the zeros in the inverse covariance matrix, also known as the precision matrix.  If the $(i,j)$th entry of the precision matrix is zero, then the $i$th and the $j$th variables are conditionally independent given all other variables;  see \cite{lauritzen_graphical_1996} and \cite{edwards_introduction_2000} for more  information on the properties of Gaussian graphical models (GGMs).  

If the variables are not normally distributed, an adjustment needs to be made if one wants to use the statistical properties of a Gaussian graphical model.  The standard procedure is to transform the data, typically by the logarithmic or the square root transformation.  With such transformations, one has to check each transformation to see if the data appear close to normal, checking which can be tedious and the method is somewhat ad-hoc.  From a modeling perspective, an easier solution is to leave the transformation functions unspecified and estimate them instead.  The nonparanormal graphical model \citep{liu_nonparanormal:_2009} consists of estimating the transformation functions using a truncated marginal empirical distribution function, and then estimating the precision matrix of the transformed variables using the graphical lasso assuming sparsity. A Bayesian approach for nonparanormal graphical models, developed by \cite{mulgrave_bayesian_2018, mulgrave_regression_2018}, uses a random series B-splines prior to estimate the transformation functions and induces sparsity on the off-diagonal entries of the precision matrix by spike-and-slab or continuous shrinkage priors, either directly or through the Cholesky decomposition. 

 However, one important requirement of the nonparanormal model is that the transformation functions need to be estimated.  In order to avoid estimating the transformation functions, alternative rank-based procedures can be employed to transform the data to normally distributed random variables.   \cite{dabrowska_partial_1988} study the properties of the likelihood function based on transformed observations in a general class called transformation models. An example is given by the rank-likelihood which is the joint distribution of the ranks of the observations. The rank-likelihood is invariant under monotone transformations of the observations.  Thus one can ignore the transformations and focus on the main parameter of interest, the precision matrix, or equivalently, the inverse correlation matrix.  The rank-likelihood has been used for semiparametric copula estimation \citep{hoff_extending_2007}, for Receiver Operating Characteristic (ROC) curve estimation \citep{gu_bayesian_2009,gu_bayesian_2014}, and for ROC surface estimation \citep{zhu_bayesian_2019}.  The use of the rank-likelihood results in a gain in robustness and simplification of estimation. Models which can be represented by transformations on Gaussian graphical models are generally known as Gaussian copula graphical models (GCGMs) and have been explored in the Bayesian literature \citep{pitt_efficient_2006, dobra_copula_2011,liu_high-dimensional_2012,mohammadi_bdgraph:_2017}. In the frequentist literature, nonparametric rank-based correlation coefficient estimators have been used to construct GCGMs \citep{liu_high-dimensional_2012,xue_regularized_2012}.  These models can also address binary or ordinal data, but we do not pursue this direction here. 

Once the transformed variables have been obtained, one needs to estimate a sparse precision matrix in order to learn the structure of the graphical model.  In the frequentist literature, the popular algorithm to use is the graphical lasso
\citep{friedman_sparse_2008}.  Numerous algorithms have been proposed to solve this problem \citep{meinshausen_high-dimensional_2006, yuan_model_2007,friedman_sparse_2008,rothman_sparse_2008,banerjee_model_2008, daspremont_first-order_2008,lu_smooth_2009, scheinberg_sparse_2010, witten_new_2011,mazumder_graphical_2012}.
Alternative penalties used to estimate the sparse precision matrix include adaptive LASSO and SCAD penalties \citep{fan_network_2009}, and LASSO and grouped LASSO penalties \citep{friedman_applications_2010}.
In the Bayesian literature, graphical models can be learned with the use of continuous shrinkage priors.  Priors such as the double exponential \citep{wang_bayesian_2012,peterson_inferring_2013}, 
uniform shrinkage priors \citep{wang_class_2013}, and spike-and-slab priors \citep{wang_scaling_2015,peterson_joint_2016,li_expectation_2017,li_bayesian_2017, mulgrave_bayesian_2018}  can characterize zero and non-zero elements of the precision matrix with continuous distributions that have a mass at zero and heavy tails.  More recently, horseshoe priors have been employed for sparse precision estimation \citep{mulgrave_regression_2018,williams_bayesian_2018}.  We estimate a sparse precision matrix using a Cholesky decomposition to naturally incorporate the positive definite matrix, a horseshoe prior to regularize the matrix, and a loss procedure to threshold the matrix.  Our methods differ from other Bayesian GCGMs which use the rank-likelihood to transform the random variables. \cite{dobra_copula_2011}
use a G-Wishart prior \citep{roverato_hyper_2002} and Markov Chain Monte Carlo (MCMC) to estimate the sparse precision matrix to construct a GCGM. \cite{mohammadi_bayesian_2017} estimate a sparse precision matrix using a G-Wishart prior and birth-and-death MCMC \citep{mohammadi_bayesian_2015}. \cite{li_expectation_2017} put a normal spike-and-slab prior on the precision matrix and use an expectation-conditional maximization algorithm for estimation.

The paper is organized as follows. In the next section, we state model and prior.  In Section 3, we review the posterior computation.  In Section 4, we review the thresholding procedure and in Section 5, we discuss our tuning procedure.  We derive a posterior consistency result in Section 6 and in Section 7, we review the results of the simulation study.  Lastly, in Section 8, we describe an application with gene expression data. 

\section{Model and Prior}
\subsection{Estimation of Transformed Variables}
Given a set of observed continuous variables $X_{i,1},\ldots,X_{i,p}$  distributed from unknown marginal distributions, there exist monotone increasing transformation functions $H_1,\ldots, H_p$, such that the distribution of the continuous transformed variables is $$Y_{i,1},\ldots, Y_{i,p} =  H_1(X_{i,1}), \ldots, H_p(X_{i,p}) \sim \N_p(\mathbf{0}, \textbf{C}) $$
where $\bm{C}$ is the correlation matrix.  We make this model identifiable by centering the transformed variables and setting the covariance matrix equal to the correlation matrix.
We wish to make inference on $\boldsymbol\Psi = \textbf{C}^{-1}$ and not on the transformation functions $H_1,\ldots, H_p$.  Let $\bm{R}$ represent the array of ranks of $\bm{X}$.  Since the transformation functions are increasing, $X_{i_{r-1},j} < X_{i_{r},j}$ implies that $Y_{i_{r-1},j} < Y_{i_{r},j}$, where $i_{r}$ is the index for the position of the $r$th smallest observation of the $j$th component of $X$, such that $i = 1, \ldots, n$ and $j = 1, \ldots, p$.  Then
for a given data set $ \bm{X} = (\bm{X}_1,\ldots, \bm{X}_p)' $, the transformed variables $ \bm{Y} = (\bm{Y}_1,\ldots, \bm{Y}_p)' $ must lie in the set
$$D=\{\bm{Y} \in \mathds{R}^{n \times p}: y_{i_{r-1},j}< y_{i_r,j}< y_{i_{r+1},j}  \}.  $$
Then following \cite{dabrowska_partial_1988} and the notation of \cite{hoff_extending_2007}, we calculate the rank-likelihood as
\begin{equation}
\textup{Pr}(\bm{Y} \in D | \bm{C}, H_1,\ldots,H_p) = \int_D p(\bm{Y}|\bm{C})d\bm{Y} = \textup{Pr}(\bm{Y} \in D|\bm{C}).
\end{equation}
Thus, this likelihood depends on the parameter $\bm{C}$ and not on the nuisance functions $H_1,\ldots, H_p$.  

Since we wish to make inference on $\boldsymbol\Psi = \textbf{C}^{-1}$, we sample $\boldsymbol\Psi$.  We reparameterize the model in terms of the non-identifiable inverse covariance matrix $\boldsymbol\Omega$, but focus our posterior inference on the identifiable inverse correlation matrix.  Thus $\boldsymbol\Psi = \bm{A}\boldsymbol\Omega \bm{A}$, where $\bm{A} = \textup{diag}(\sigma_{dd})$ and $\sigma_{dd}$ are the square roots of the diagonal elements of $\boldsymbol\Sigma = \boldsymbol\Omega^{-1}$.  The rank-likelihood is scale invariant so the non-identifiable and identifiable models lead to the same posterior distribution,  $\textup{Pr}(\boldsymbol\Psi|\bm{Y} \in D) \propto p(\boldsymbol\Psi) \times p(\bm{Y}\in D|\boldsymbol\Psi)  $.  

\subsection{Estimation of Inverse Correlation Matrix}
We put a horseshoe prior on $\boldsymbol\Omega$ and sample $\boldsymbol\Omega$ using a regression-based Cholesky decomposition method discussed in \cite{mulgrave_regression_2018}.  Denote the Cholesky decomposition of $\boldsymbol\Omega$ as $\boldsymbol\Omega = \boldsymbol L \boldsymbol L^{T}$, where $\boldsymbol L=(\!(l_{kd})\!)$ is a lower triangular matrix.  Define the coefficients $\beta_{kd} = -l_{kd}/l_{dd}$ and the precision as $\phi_d = 1/\sigma_d^2 = l_{dd}^2$. Then the multivariate Gaussian model $\bm{Y} \sim \N_p(\mathbf{0}, \boldsymbol\Sigma$), where $\boldsymbol\Sigma = \boldsymbol\Omega^{-1},$ leads to the set of independent regression problems, 
$${Y}_d = \sum_{k>d}\beta_{kd}{Y}_k + \epsilon_d, \; \epsilon_d \sim \N(0, \sigma_d^2), \; d = 1,\ldots, p,   $$
where $\beta_{kd}$ are the regression coefficients for $k = d+1, \ldots, p$ and $d = 1, \ldots, p$. Denoting $\bm{Y}_d$ to be the $d$th column of $\bm{Y}$, $\bm{Y}_{k>d}$ the matrix formed by columns of $\bm{Y}$ greater than $d$, and 
$\boldsymbol\beta_{k>d} = (\beta_{d+1},\ldots, \beta_{p})$, we may write the regression relation in the vector form as 
$$\bm{Y}_d|(\bm{Y}_{k>d}, \boldsymbol\beta_{k>d},\sigma_d^2) \sim \N(\bm{Y}_{k>d}\boldsymbol\beta_{k>d}, \sigma_d^2\bm{I}),$$
which gives rise to the likelihood. 

We use a standard conjugate noninformative prior on the variances with improper density proportional to $\sigma_d^{-2}$.  We enforce a sparsity constraint along the rows of the lower triangular matrix in order to ensure that the probability that an entry is nonzero (i.e. sparsity) remains roughly the same over different rows.  We choose $\rho_k = $\textup{Prob}(nonzero in $k$th row)$ ={c}/({p\sqrt{k}})$, and tune the value of $c \in \{0.1,1,10,100\}  $ to cover a range of four orders of magnitude, i.e. $10^{-1}, \; 10^{0}, \; 10^{1}, \; 10^{2}$;  see \cite{mulgrave_regression_2018} for the more information on the sparsity constraint.

We use a horseshoe prior on $\beta_{kd}$ described in \cite{neville_mean_2014}  
\begin{equation}
\label{HorseshoePrior}
\begin{aligned}
\beta_{kd} |( \lambda_d^2, b_{kd}, \sigma_d^2 ) \stackrel{\mathrm{ind}}{\sim} \N(0, \frac{\sigma_d^2 b_{kd} c^2 \lambda_d^2}{p^2 k} ),\\
 \lambda_d^2|a_d \sim \mathrm{IG}(\frac{1}{2}, \frac{1}{a_d}), \\
a_d \sim \mathrm{IG}(\frac{1}{2}, 1), \\
b_{kd}|h_{kd} \stackrel{\mathrm{ind}}{\sim} \mathrm{IG}(\frac{1}{2}, \frac{1}{h_{kd}}),\\
h_{kd} \sim \mathrm{IG}(\frac{1}{2},1), \\
\sigma_d^2 \sim \mathrm{IG}(0.01,0.01).  
\end{aligned}
\end{equation}
for $k,d = 1,\ldots,p$; here IG stands for the inverse gamma distribution.

According to  \cite{van_der_pas_horseshoe_2014}, the global scale parameter $\lambda$ is roughly equivalent with the probability of a nonzero element.  We enforce the sparsity constraint by replacing $\lambda$ with $(\lambda_d c)/(p\sqrt{k})$.  Thus, since we are working with the squared parameter, the factor in the variance term for $\beta_{kd}$ is $(\lambda^2 c^2)/(p^2 k)$.  
The prior on $\boldsymbol\Omega$ leads to an induced prior on $\boldsymbol\Psi$.

\section{Posterior Computation} 
We obtain samples of $\textup{Pr}(\boldsymbol\Psi | \bm{Y} \in D)  $ by employing the following Gibbs sampler: 
\begin{enumerate}
\item Sample $\bm{Y} \sim \N(\mathbf{0}, \boldsymbol\Psi^{-1}):$

For $d = 1,\ldots,p$,
\begin{enumerate}[(a)]
\item Compute the ranks of $\bm{X}_d$, where $r = 1, \ldots, n$ represents the ranks.  

For $r = 1, \ldots, n,$ 
\begin{enumerate}[(i)]
\item Compute $y_{\textup{lower}} = y_{i_{r-1},d}$ and  $y_{\textup{upper}}  = y_{i_{r+1},d}$, where $i$ is the index for the position of the $r$-th rank of $\bm{X}_d$.  Let $y_{i_{0},d} = -\infty$ and $y_{i_{n+1},d} = \infty$.

\item Compute $\mu_{i_{r},d} = -\boldsymbol\psi_{d,d}^{-1}\boldsymbol\Psi_{d,\setminus d}\bm{Y}_{i_{r},\setminus d}  $
\item $\sigma_d^2 = \boldsymbol\psi_{d,d}^{-1}$
\item Sample $Y_{i_{r},d} \sim \mathrm{TN}(\mu_{i_{r},d}, \sigma_d^2;\; y_{\textup{lower}} < Y_{i_{r},d}< y_{\textup{upper}}),$ where \\ $\mathrm{TN}(a, b;\; c<x<d)$ denotes a univariate truncated normal distribution with mean $a$, variance $b$, and truncation limits $c$ and $d$.  Sampling from the truncated normal distribution is implemented with the fast sampling function {\tt trandn} in {\tt MATLAB} that uses minimax tilting \citep{botev_normal_2017}.

\end{enumerate}
\end{enumerate}
\item Sample $\boldsymbol\Omega$:

For $d = 1, \ldots, p-1,$

\begin{enumerate}[(a)]
\item  Sample the variables 
$\boldsymbol\beta_{k>d}|(\sigma_d, \bm{b}_{k>d}, \lambda_d^2)  \sim \N(\bm{A}^{-1}\bm{Y}_{k>d}^T \bm{Y}_d, \sigma_d^2\bm{A}^{-1})$, where \\
$\bm{A} = (\bm{Y}_{k>d}'\bm{Y}_{k>d} + \textup{diag}({p^2 k}/({\lambda_d^2 \bm{b}_{k>d} c^2}))).$

Since sampling from this normal distribution  can be expensive with large $p$, we used an exact sampling algorithm for Gaussian priors that uses data augmentation \citep{bhattacharya_fast_2016}:

\begin{enumerate}[(i)]
\item Sample $t \sim \N(\mathbf{0}, \bm{D})$ and $\delta \sim \N(0, I_n)$, where $\bm{D} = \displaystyle \sigma_d^2\textup{diag}(\frac{\lambda_d^2\bm{b}_{k>d} c^2}{p^2 k});$
\item set $v = \boldsymbol\Phi t + \delta$, where $\boldsymbol\Phi = \bm{Y}_{k>d}/\sigma_d;$
\item solve $ (\boldsymbol\Phi \bm{D}\boldsymbol\Phi' + \bm{I}_n)w = (\alpha - v)$, where $\alpha =  \textbf{Y}_d/\sigma_d;$
\item set $\beta = t + \bm{D} \boldsymbol\Phi'w. $
\end{enumerate}

\item Sample $
\lambda_d^2 \sim \mathrm{IG}(\displaystyle \frac{\#(k>d)}{2} + \frac{1}{2}, \frac{1}{2}\boldsymbol\beta_{k>d}'\textup{diag}(\frac{p^2 k}{\sigma_d^2 \bm{b}_{k>d} c^2 })\boldsymbol\beta_{k>d} + \frac{1}{a_d}).$
\item Sample $a_d \sim \mathrm{IG}(1, \lambda_d^{-2} +1).$
\item Sample $b_{kd} \sim  \mathrm{IG}(1, \frac{p^2 k \beta_{kd}^2}{2 \sigma_d^2\lambda_d^2 c^2} + \frac{1}{h_{kd}}).$
\item Sample $h_{kd} \sim \mathrm{IG}(1, \frac{1}{b_{kd}} + 1).$
\item Sample $
\sigma_d^2 \sim  \mathrm{IG}( \frac{n + \#(k>d)}{2} + 0.01,
\frac{1}{2}\left\Vert \bm{Y}_d - \bm{Y}_{k>d}\boldsymbol\beta_{k>d}\right \Vert^2$  \\ 
$+ \frac{1}{2}\boldsymbol\beta_{k>d}'\textup{diag}(\frac{p^2 k}{\lambda_d^2 \bm{b}_{k>d} c^2})\boldsymbol\beta_{k>d} + 0.01).
$
Sample 
$\sigma_p^2 \sim \mathrm{IG}({n}/{2} + 0.01, 1/2\left \Vert \bm{Y}_p\right \Vert^2 + 0.01).$
\item Compute
$l_{kd} = -\beta_{kd}/\sigma_d \textup{ and } l_{dd} = 1/\sigma_d.$

\item Compute
$\boldsymbol\Omega = \textbf{L}\textbf{L}'.$
\end{enumerate}

\item Set $\boldsymbol\Psi = \bm{A}\boldsymbol\Omega \bm{A}$, where $\bm{A} = \textup{diag}(\sigma_{dd})$ and $\sigma_{dd}$ are the square roots of the diagonal elements of $\boldsymbol\Sigma = \boldsymbol\Omega^{-1}$. 
\end{enumerate}
These steps are repeated until convergence.

 \section{Thresholding}

\subsection{0-1 Loss Procedure}
\label{loss}
We find the posterior partial correlation using the inverse correlation matrices from the Gibbs sampler of the horseshoe prior \eqref{HorseshoePrior} and the posterior partial correlation using the standard conjugate Wishart prior.  The posterior partial correlation using the matrices from the Gibbs sampler is defined as 
\begin{equation*}
\rho_{kd,m} = -\frac{\psi_{kd,m}}{\sqrt{\psi_{kd,m}\psi_{dd,m}}},
\end{equation*}
where $\psi_{kd,m}$ is the $m$th sample of $M$ Markov chain Monte Carlo (MCMC) draws after burn-in from the posterior distribution of $\psi_{kd}$, $k,d = 1,\ldots,p$, $m = 1,\ldots, M$.  The posterior partial correlation using the standard conjugate Wishart prior is found by starting with the latent observation, $\bm{Y}_m,$ which is obtained from the MCMC output.  We put a standard Wishart prior on the inverse correlation matrix, $\boldsymbol\Psi \sim \mathrm{W}_p(3, \bm{I})$, where $\bm{I}$ is the identity matrix.  By conjugacy, the posterior is $\mathrm{W}_p(n+3, (\bm{I} + \bm{S})^{-1})$, where $\bm{S} = \bm{Y}'\bm{Y}$.  We compute  the mean of the posterior distribution given $\bm{Y}$, $\boldsymbol\Lambda = \mathds{E}(\boldsymbol\Psi|\bm{Y}) = (n+3)(\bm{I}_p + \bm{S})^{-1}$.  Finally, we find the posterior partial correlation coefficients 
\begin{equation*}
\phi_{kd,m} = -\frac{\lambda_{kd,m}}{\sqrt{\lambda_{kd,m}\lambda_{dd,m}}}, 
\end{equation*}
where $\lambda_{kd,m}$ is the $(k,d)$th entry of $\boldsymbol\Lambda$ at the $m$th MCMC iteration. 

We link these two posterior partial correlations for the 0-1 loss method.  Our convention is that the event $\{\psi_{kd,m} \ne 0\}$ holds if and only if 
\begin{equation}
\frac{\rho_{kd,m}}{\phi_{kd,m}} > 0.5
\end{equation}
for $k,d = 1,\ldots,p$ and $m = 1,\ldots, M$.  The rationale for this thresholding procedure is that we are comparing the regularized precision matrix to the non-regularized precision matrix from the Wishart prior.  If the absolute value of the partial correlation coefficient from the regularized precision matrix is similar in size or larger than the absolute value of the partial correlation coefficient from the Wishart precision matrix, then the edge matrix should have an edge.  If the absolute value of the partial correlation coefficient from the regularized precision matrix is much smaller than the absolute value of the coefficient from the Wishart matrix, then the edge matrix should not have an edge.  The precision matrix from the Wishart prior serves as a means of comparison to determine whether the element of the regularized precision matrix is truly large or small.

\section{Choice of Prior Parameters}
For the sparsity constraint on the inverse correlation matrix $\boldsymbol\Psi$, we need to select the value of the parameter $c$.  We solve a convex constrained optimization problem in order to use the Bayesian Information Criterion (BIC), originally described in \cite{dahl_maximum_2005, dahl_covariance_2008} and developed further in \cite{mulgrave_bayesian_2018}.  First, we find the Bayes estimate of the  inverse correlation matrix, $\hat{\boldsymbol\Psi} = \mathrm{E}(\boldsymbol\Psi | \bm{Y})$.  We also find the average of the transformed variables, $\bar{\bm{Y}} = M^{-1}\sum_{m = 1}^M\bm{Y}_m$, where $\bm{Y}_m$, $m = 1, \ldots, M$, are obtained from the MCMC output.  Then, using the sum of squares matrix, $\bm{S} =\bar{\bm{Y}}'\bar{\bm{Y}}$, we solve the following to obtain the maximum likelihood estimate of the inverse correlation matrix, $\hat{\boldsymbol\Psi}_{\mathrm{MLE}}$:
\begin{equation*}
\underset{\boldsymbol\Psi}{\text{minimize }} 
-n\log\det \boldsymbol\Psi + \textup{tr}(\boldsymbol\Psi \textbf{S}), \quad 
 \text{subject to }
\mathcal{C}({\boldsymbol\Psi}),
\end{equation*}
where $\mathcal{C}$ stands for the constraint that an element of $\bm{\Psi}$ is zero if and only if  the estimated edge matrix from the MCMC sampler has that element as zero.  The estimated edge matrix from the MCMC sampler will be described in more detail in Subsection \ref{performanceassessment}.  For computational simplicity, in the code, we represent this problem as an unconstrained optimization problem as described in \cite{dahl_maximum_2005,dahl_covariance_2008}.  

Lastly, we calculate 
$\textup{BIC} = -2\ell(\hat{\boldsymbol\Psi}_{\mathrm{MLE}}) + k\log n  $, 
where $k = \#\mathcal{C}(\hat{\boldsymbol\Psi}_{\mathrm{MLE}})$, the sum of the number of diagonal elements and the number of edges in the estimated edge matrix, and $-\ell(\hat{\boldsymbol\Psi}_{\mathrm{MLE}}) = -n\log \det \hat{\boldsymbol\Psi}_{\mathrm{MLE}} + \textup{tr}(\hat{\boldsymbol\Psi}_{\mathrm{MLE}}\textbf{S})$. 
We select the $c$ that results in the smallest BIC.

\section{Posterior Consistency}

In this section, we show that in the fixed dimensional setting (i.e. $p$ is fixed), the rank-based posterior distribution of $\bm{\Psi}$ is consistent at its true value $\bm{\Psi}_0$ for almost all $\bm{\Psi}_0$ with respect to the Lebesgue measure under the only assumption that the prior distribution for $\bm{\Psi}$ has positive density on the space of inverse correlation matrices, i.e., the collection of all $p\times p$ positive definite matrices $\bm{\Psi}$ such that all diagonal elements of $\bm{\Psi}^{-1}$ are $1$. 

\begin{theorem} 
\label{thm:posterior consistency}
Assume that $\boldsymbol\Psi$ has prior density $\pi(\boldsymbol\Psi) > 0$ a.e. over the space of positive definite matrices, with respect to the Lebesgue measure $\nu$ and that the dimension $p$ is fixed.  Then for $\boldsymbol\Psi_0$ a.e. $[\nu]$, and for any neighborhood $\mathcal{U}_0$ of $\boldsymbol\Psi_0$, we have that 
\begin{equation}
\label{eq:consistency}
\lim_{n\to\infty} \pi(\boldsymbol\Psi \in \mathcal{U}_0 | \bm{R}) = 1 \mbox{ a.e. } [P^{\infty}_{\boldsymbol\Psi, H}],
\end{equation}
where $[P^{\infty}_{\boldsymbol\Psi, H}]$ denotes the joint distribution of all $\bm{X}$'s and ranks $\bm{R}$ with $\boldsymbol\Psi_0$ as the true value of $\boldsymbol\Psi$ and $\bm{H}$ denotes the underlying normality restoring transformations.
\end{theorem}

\begin{proof}
The proof is based on an application of Doob's Theorem \citep{ghosal_fundamentals_2017}, Section~6.2. Doob's theorem is a very general posterior consistency result, which only requires that in the joint distribution of all parameters and observables, the parameter can be a.s. written as a function of all observables of all stages, and then concludes that posterior consistency holds for almost all parameters with respect to the prior distribution. Here observations are the rank information $\bm{R}_{n}=(\!(R_{n,ik})\!)$ for each variable, where $R_{n,ik}$ stands for the rank of the $i$th observation in the $k$th component, $i=1,\ldots,n$, $k=1,\ldots,p$, $n=1,2,\ldots$. 
We follow some arguments given in \cite{gu_bayesian_2009}. Let $U_{ik}=F_k(X_{ik})$, the ``population quantile'' of the $i$th observation regarding the $k$th variable, $i=1,\ldots,n$, $k=1,\ldots,p$. Note that $Y_{ik}=\Phi^{-1}(U_{ik})$. By Theorem a on page 157 of \cite{hajek_theory_1967}, for any $k=1,\ldots,p$,  
\begin{equation}
\mathrm{E}(U_{ik} - \frac{R_{n,ik}}{n+1})^2 = \frac{1}{n}\sum_{j=1}^n \mathrm{E}[(U_{i,k} - \frac{j}{n+1})^2|R_{n,ik} = j ] = \frac{1}{n}\sum_{j=1}^n\frac{j(n-j+1)}{(n+1)^2(n+2)} < \frac{1}{n}.
 \end{equation}
 As such, $U_{ik}$ is an in-probability limit of $\mathcal{F}_n$-measurable random variables, where  $\mathcal{F}_n$ is the $\sigma$-field generated by $\{R_{n,ik}: i=1,\ldots,n,\, k=1,\ldots,p\}$.  Thus, for any $k=1,\ldots,p$, 
 \begin{equation}
 U_{ik} = \lim_{n' \to \infty} \frac{R_{n',ik}}{n' + 1} \mbox{ for } i \geq 1, \mbox{ with probability 1 for some subsequence } \{n'\}
 \end{equation}
 and hence, $U_{ik}$ is an $\mathcal{F}_{\infty}$-measurable random variable, where $\mathcal{F}_{\infty} = \sigma \langle \cup_{1}^{\infty}\mathcal{F}_n \rangle$, and so is $Y_{ik}=\Phi^{-1}(U_{ik})$. Therefore it suffices to show that $\bm{\Psi}$ can be written as the almost sure limit of a sequence of functions of $\{ Y_{ik}: i=1,\ldots,n, \, k=1,\ldots,p\}$. 
 
Let $C_{jk}$ stand for the $(j,k)$th element of $\bm{C}=\bm{\Psi}^{-1}$. Then clearly $C_{jk}=n^{-1} \lim_{n\to\infty} \sum_{i=1}^n Y_{ij} Y_{ik}$ almost surely. Thus $\bm{C}$, and hence $\bm{\Psi}$ is expressible as an almost sure limit of $\{ Y_{ik}: i=1,\ldots,n, \, k=1,\ldots,p\}$. 
 Thus, by Doob's theorem, the consistency of the posterior \eqref{eq:consistency}  at $\boldsymbol\Psi_0$ holds a.e. $[\pi]$. However, as the prior density is positive throughout the parameter space, it also follows that the posterior \eqref{eq:consistency} at $\boldsymbol\Psi_0$ holds a.e. $[\nu]$. 
\end{proof}

Usually, the main criticism against a posterior consistency result obtained by applying Doob's theorem is that the exceptional set where consistency may fail may be ``large'' since it only needs to be null with respect to the prior, which is somewhat arbitrary. However, in the present application, since the parameter space for the parameter of interest $\bm{\Psi}$ is finite dimensional, where we have the Lebesgue measure as a benchmark measure, the exceptional set of points where the posterior may be inconsistent is characterized as Lebesgue null, which can be regarded as ``small''. It is important to note that the normality restoring transformations are taken to be fixed, and no prior is assigned on them. Since the underlying procedure is unaffected by the transformations, the fact that these transformations are unknown does not matter. Note that the fixed dimensionality of the variables is essential since the posterior consistency result is applicable only if the parameter space is fixed. 

\section{Simulation Results}
\label{simulationresults}
We conduct a simulation study to compare the performance of the proposed Bayesian rank-likelihood method, the Bayesian GCGM \citep{mohammadi_bayesian_2017}, the empirical method in a nonparanormal graphical model \citep{liu_nonparanormal:_2009},  the Bayesian method in a nonparanormal graphical model \citep{mulgrave_regression_2018}, and the empirical method based on GCGM \citep{liu_high-dimensional_2012}.  Both the proposed Bayesian rank-likelihood method, indicated as Rank-Likelihood, and the Bayesian method of \cite{mulgrave_regression_2018}, indicated as B-splines, use a horseshoe prior on the Cholesky decomposition of the precision matrix and MCMC estimation. The B-splines method uses a random series B-splines prior to estimate the transformation functions. The Bayesian method based on GCGM, indicated as Bayesian Copula, uses the rank-likelihood to transform the random variables and puts a G-Wishart prior on the inverse correlation matrix, a uniform prior on the graph, and estimates the sparse matrix using a birth-and-death MCMC \citep{mohammadi_bayesian_2015}.  The empirical method in a nonparanormal graphical model, indicated as Truncation, uses a truncated marginal empirical distribution function to transform the variables and the graphical lasso to estimate the sparse precision matrix.  The empirical method based on GCGM, indicated as SKEPTIC, uses Spearman's rho to transform the variables and estimates the sparse precision matrix with the graphical lasso.

 The random variables, $Y_1,\ldots,Y_p$, are simulated from a multivariate normal distribution such that $Y_{i1},\ldots,Y_{ip} \stackrel{\mathrm{i.i.d.}}{\sim} \N(\boldsymbol\mu, \boldsymbol\Omega^{-1})$ for $i = 1, \ldots, n$.  The means $\boldsymbol\mu$ are selected from an equally spaced grid between 0 and 5 with length $p$.  We consider nine different combinations of $n, p,$ and sparsity for $\boldsymbol\Omega$:  
\begin{itemize}
\item $p=25,\; n=50$, AR(4) model;
\item $p=50$, $n=100$, AR(4) model;
\item $p=100$, $n=500$, AR(4) model;
\item $p=25,\; n=50$, AR(1) model;
\item $p=50$, $n=100$, AR(1) model;
\item $p=100$, $n=500$, AR(1) model;
\item $p=25$, $n=50$, sparsity = $10\%$ non-zero entries in the off-diagonals;
\item $p=50$, $n=100$, sparsity = $5\%$ non-zero entries in the off-diagonals;
\item $p=100$, $n=500$, sparsity = $2\%$ non-zero entries in the off-diagonals;
\end{itemize}
 where the AR(4), AR(1), and star models are described by the relations 
\begin{itemize} 
\item AR(4) model: $\omega_{i,i} = 1, \; \omega_{i, i-1} = \omega_{i-1,i} =  0.2, \; \omega_{i, i-2} = \omega_{i-2,i} =  0.2, \; \omega_{i, i-3} = \omega_{i-3,i} =  0.2, \; \omega_{i, i-4} = \omega_{i-4,i} =  0.1$;
\item AR(1) model: $\omega_{1,1} = 1.9608, \; \omega_{i, i-1} = \omega_{i-1,i} =  -1.3725$, $\omega_{p, p} = 1.9608  $
\end{itemize}

The percent sparsity levels for $\boldsymbol\Omega$ are computed using lower triangular matrices that have diagonal entries normally
distributed with mean 1 and standard deviation 0.1, and 
non-zero off-diagonal entries normally distributed
with mean 0 and standard deviation 1.  

The observed variables $\bm{X} = (X_1, \ldots,X_p)$ are constructed from the simulated variables $Y_1, \ldots, Y_p$.  The functions used to construct the observed variables were four cumulative distribution functions (c.d.f.s): asymmetric Laplace, extreme value, logistic, and stable.  We could choose any values of the parameters for the c.d.f.s, but instead of selecting them ourselves, we automatically choose the values of the parameters to be the maximum likelihood estimates with the {\tt mle} function in {\tt MATLAB}.  The values of the parameters for each of the c.d.f.s are the maximum likelihood estimates for the parameters of the corresponding distributions (asymmetric Laplace, extreme value, logistic, and stable), using the variables $Y_1, \ldots, Y_p$.

We follow the procedure in \cite{mulgrave_bayesian_2018} to estimate the transformation functions for the B-splines method.  The hyperparameters for the normal prior are chosen to be $\nu = 1, \tau = 1,$ and $\sigma^2 = 1$.   To choose the number of basis functions, we use the Aikaike Information Criterion.  Samples from the truncated multivariate normal posterior distributions for the B-spline coefficients are obtained using the exact Hamiltonian Monte Carlo (exact HMC) algorithm \citep{pakman_exact_2014}.  
After finding the initial coefficient values $\boldsymbol\theta_d$, we construct initial values for $Y_{d, \textup{initial}} =  \sum_{j=1}^J\theta_{dj, \textup{initial}}B_j(X_{d})$ using the observed variables.  These initial values  $\bm{Y}_{\textup{initial}}$ are used to find initial values for $\boldsymbol\Sigma, \boldsymbol\mu$, and $\boldsymbol\Omega$ for the algorithm.

For the Rank-Likelihood method, we initialize the algorithm  using the ranks of the observed variables.  First, we create a $p\times n$ matrix of ranks, $\bm{R} = \mbox{rank}(\bm{X})$.  Then, we divide the columns by $n+1$ and 
transform each entry by the standard normal quantile function using the inverse standard normal c.d.f., such that $\bm{Y}_{\textup{initial}} = \boldsymbol\Phi^{-1}(\bm{R}/(n+1))'$.  Finally, we take the transpose to obtain an $n\times p$ matrix.  We obtain an initial value for the inverse correlation matrix using $\boldsymbol\Psi_{\textup{initial}} = (\mbox{corr}(\bm{Y}_{\textup{initial}}))^{-1},$ where $\mbox{corr}$ stands for the correlation coefficients of $\bm{Y}_{\textup{initial}}.$

For both the Rank-Likelihood and the B-splines methods, the hyperparameters for the horseshoe prior are initialized with ones. To impose the sparsity constraint on the Cholesky decomposition of the matrices for the B-splines and the Rank-Likelihood methods, we consider four values for tuning: $c \in \{0.1,1,10,100\}$.  We select the graphical model with the value of $c$ having the lowest BIC. The 0-1 loss procedure \eqref{loss} is used to threshold the matrices for the B-splines and the Rank-Likelihood methods and construct the corresponding edge matrices.  The codes for these methods are written in {\tt MATLAB}.

For the simulation study, we run 100 replications for each of the nine combinations and assess structure learning and parameter estimation for each replication.  We collect $10000$ MCMC samples for inference after discarding a burn-in of $5000$ and we do not apply thinning.  The Bayesian Copula method is implemented in the {\tt R} package {\tt BDgraph} \citep{mohammadi_bdgraph:_2017, mohammadi_bdgraph:_2019} using the option ``gcgm''. Bayesian model averaging is used for inverse correlation matrix and graph selection. The default option in the {\tt BDgraph} package selects the graph formed by links having
estimated posterior probabilities greater than $0.5$.  The Truncation and SKEPTIC methods are implemented in the {\tt R} package {\tt huge}  \citep{zhao_huge:_2015} using the ``truncation'' and ``skeptic'' options respectively.  For both empirical methods, the graphical lasso method is used for the graph estimation based on transformed variables and the default lossless screening method \citep{witten_new_2011,mazumder_exact_2012} is applied.  A sequence of 100 regularization parameters, $\lambda$, is generated starting from $\lambda_{\mbox{max}}$ to $0.01*\lambda_{\mbox{max}}$, in the log-scale.  We define $\lambda_{\mbox{max}} = \max(\max(\bm{M} - \mbox{diag}(p)), -\min(\bm{M} - \mbox{diag}(p)))$, where $\bm{M}$ is the correlation matrix from the SKEPTIC method and $\bm{M}$ is the matrix constructed for the Truncation method after scaling the transformed variables and converting it to a correlation matrix.  Note that $p$ is the dimension.  The method used to select the graphical model along the regularization path is Generalized Stability Approach to
Regularization Selection \citep{muller_generalized_2016}, or G-StARS, implemented using the {\tt R} package {\tt pulsar}.  We use upper and lower bounds to reduce the computational burden and the number of random subsamples taken for graph re-estimation was 100.  All codes are provided in the Supplementary Material.

\subsection{Performance Assessment}
\label{performanceassessment}
For the Rank-Likelihood method, we find the Bayes estimate of the inverse correlation matrix $\hat{\boldsymbol\Upsilon} = \mathds{E}(\boldsymbol\Psi|\bm{Y})$ and average it over MCMC iterations. For the B-splines method, the Bayes estimate of the precision matrix is $\hat{\boldsymbol\Upsilon} = \mathds{E}(\boldsymbol\Omega|\bm{X})$, using the MCMC samples.  The median probability model \citep{berger_optimal_2004} is used to find the Bayes estimate of the edge matrix for both the Rank-Likelihood and B-splines methods. We find the estimated edge matrix by first using the 0-1 loss procedure \eqref{loss} to threshold the MCMC inverse correlation and precision matrices, and then we take the mean of the thresholded matrices.  If the off-diagonal element of the mean is greater than $0.5$, the element is registered as an edge; else it is registered as a no-edge.

We compute the specificity (SP), sensitivity (SE), and Matthews Correlation Coefficient (MCC), previously used for assessing the accuracy of classification procedures \citep{baldi_assessing_2000}, to assess the performance of the graphical structure learning.  They are defined as follows:
\begin{align*}
\textup{Specificity} = \frac{\textup{TN}}{\textup{TN} + \textup{FP}}, \qquad \textup{Sensitivity} = \frac{\textup{TP}}{\textup{TP} + \textup{FN}},\\ 
\\
\textup{MCC} = \frac{\textup{TP} \times \textup{TN} - \textup{FP} \times \textup{FN}}{\sqrt{(\textup{TP} + \textup{FP})(\textup{TP} + \textup{FN})(\textup{TN} + \textup{FP})(\textup{TN} + \textup{FN})}},
\end{align*}
where TP stands for true positives, TN stands for true negatives, FP stands for false positives, and FN stands for false negatives.  Specificity and sensitivity values are between 0 and 1, where 1 is the best value.  MCC values are between $-1$ and 1, and 1 is the best value.  

We also assess the strength of parameter estimation.  We consider the scaled $L_1$-loss function, the average absolute distance.  The scaled $L_1$-loss is defined as 
$$\textup{Scaled $L_1$-loss} = \frac{1}{p^2}\sum_k\sum_d \left \| \hat{\boldsymbol\Upsilon}_{kd} -\boldsymbol\Upsilon_{\textup{true}_{kd}} \right \|.
$$
For the Rank-Likelihood, SKEPTIC, and Bayesian Copula methods, $\hat{\boldsymbol\Upsilon}$ is the estimated inverse correlation matrix and $\boldsymbol\Upsilon_{\textup{true}}$ is the true inverse correlation matrix.  For the Truncation and B-splines method, $\hat{\boldsymbol\Upsilon}$ is the estimated inverse covariance matrix and $\boldsymbol\Omega_{\textup{true}}$ is the true inverse covariance matrix.  The results are presented in Figures 1--4.

The Rank-Likelihood method performs consistently better than the B-splines method in terms of structure learning and parameter estimation.  In particular, the Rank-Likelihood method  appears to be more sensitive and specific to signals than the B-splines method.   In addition, the scaled $L_1$-loss of the Rank-Likelihood method is significantly better than the scaled $L_1$-loss of the B-splines method.  The Bayesian Copula method performs similar to or better than SKEPTIC and Truncation methods in terms of structure learning for all models considered.  The Bayesian Copula method performs similar to the SKEPTIC and Truncation methods with regard to parameter estimation. The proposed Rank-Likelihood method performs the best at structure learning for all models considered except for the AR(4) model at dimension $p=100$, at which the Bayesian Copula model performs the best. For parameter estimation, the proposed Rank-Likelihood method generally outperforms all competing methods.

Thus overall, compared to competing methods, the Rank-Likelihood method performs nearly the same or better for structure learning and parameter estimation for all models excluding the AR(4) model at dimension $p=100$.   However, the Rank-Likelihood model has good performance when considering structure learning and parameter estimation together, compared to the competing models.

\begin{figure}[!htbp]
    \centering
    \includegraphics[width=.9\linewidth]{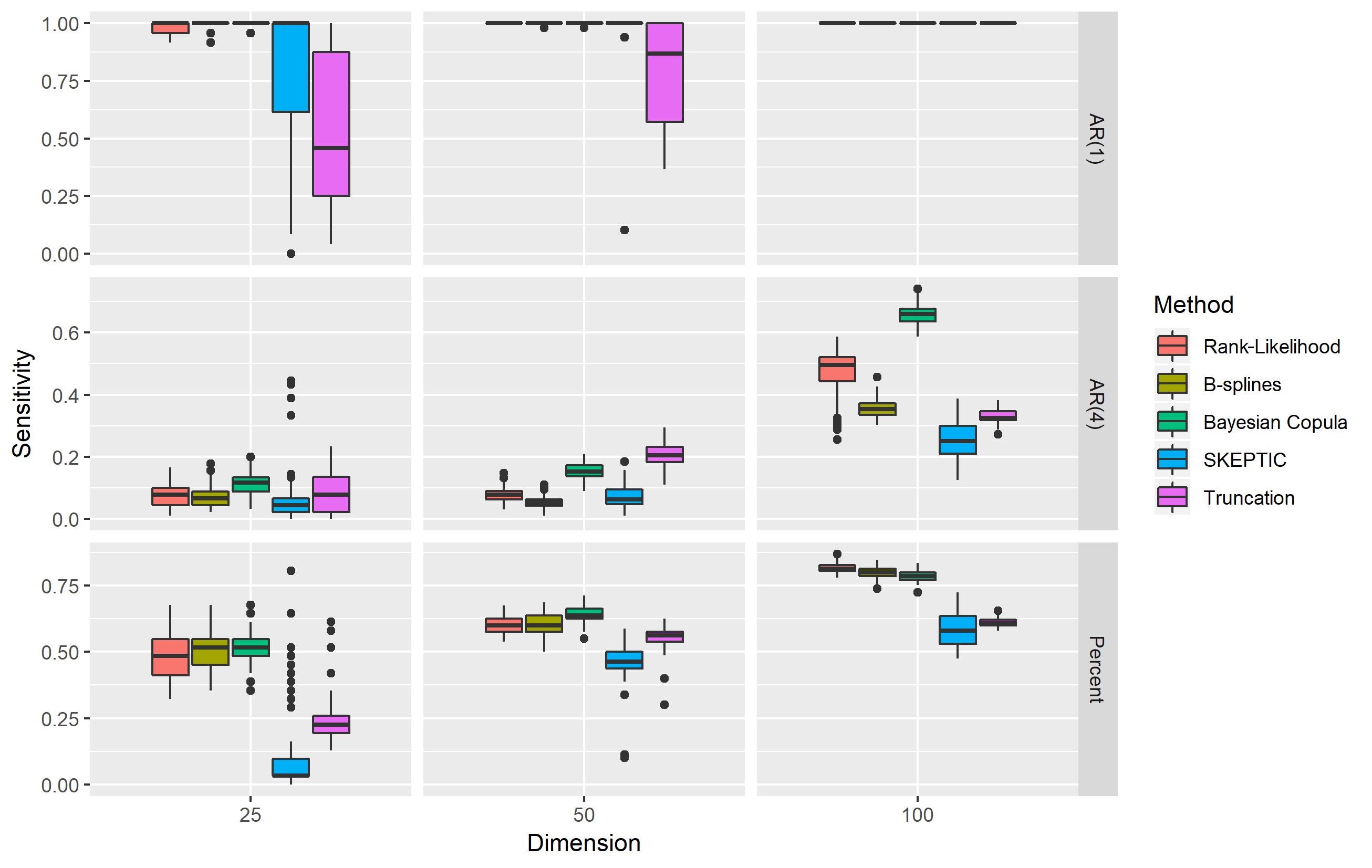}
    \caption{Boxplots of the sensitivity results for each of the methods for different structures of precision matrices. Percent refers to the 10\% model for dimension $p=25$, 5\% model for dimension $p=50$ and 2\% model for dimension $p=100$.}
   \vspace{1ex}
  \end{figure}
  
  \begin{figure}[!htbp]
    \centering
    \includegraphics[width=.9\linewidth]{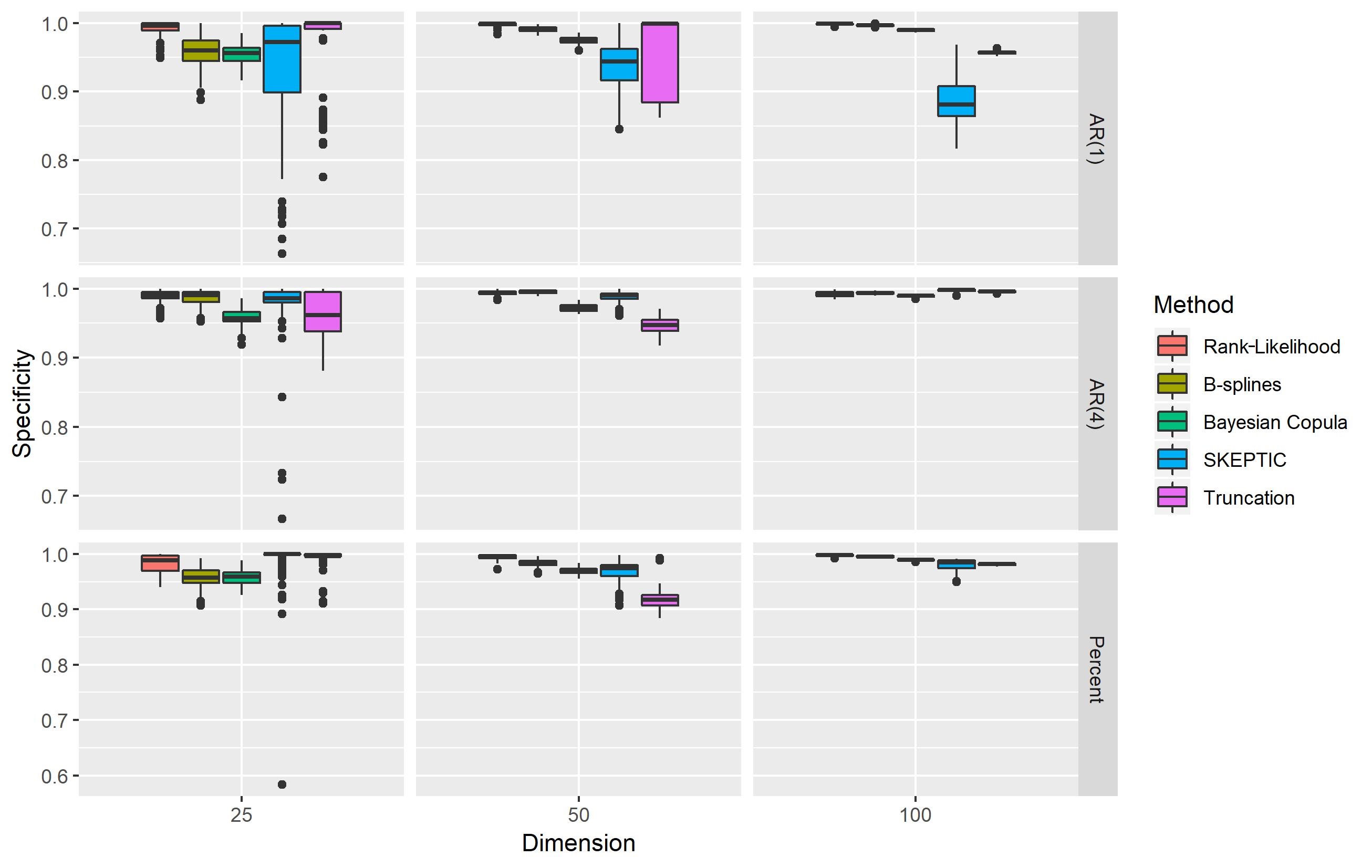}
    \caption{Boxplots of the specificity results for each of the methods for different structures of precision matrices. Percent refers to the 10\% model for dimension $p=25$, 5\% model for dimension $p=50$ and 2\% model for dimension $p=100$.}
   \vspace{1ex}
  \end{figure}
  
    \begin{figure}[!htbp]
    \centering
    \includegraphics[width=.9\linewidth]{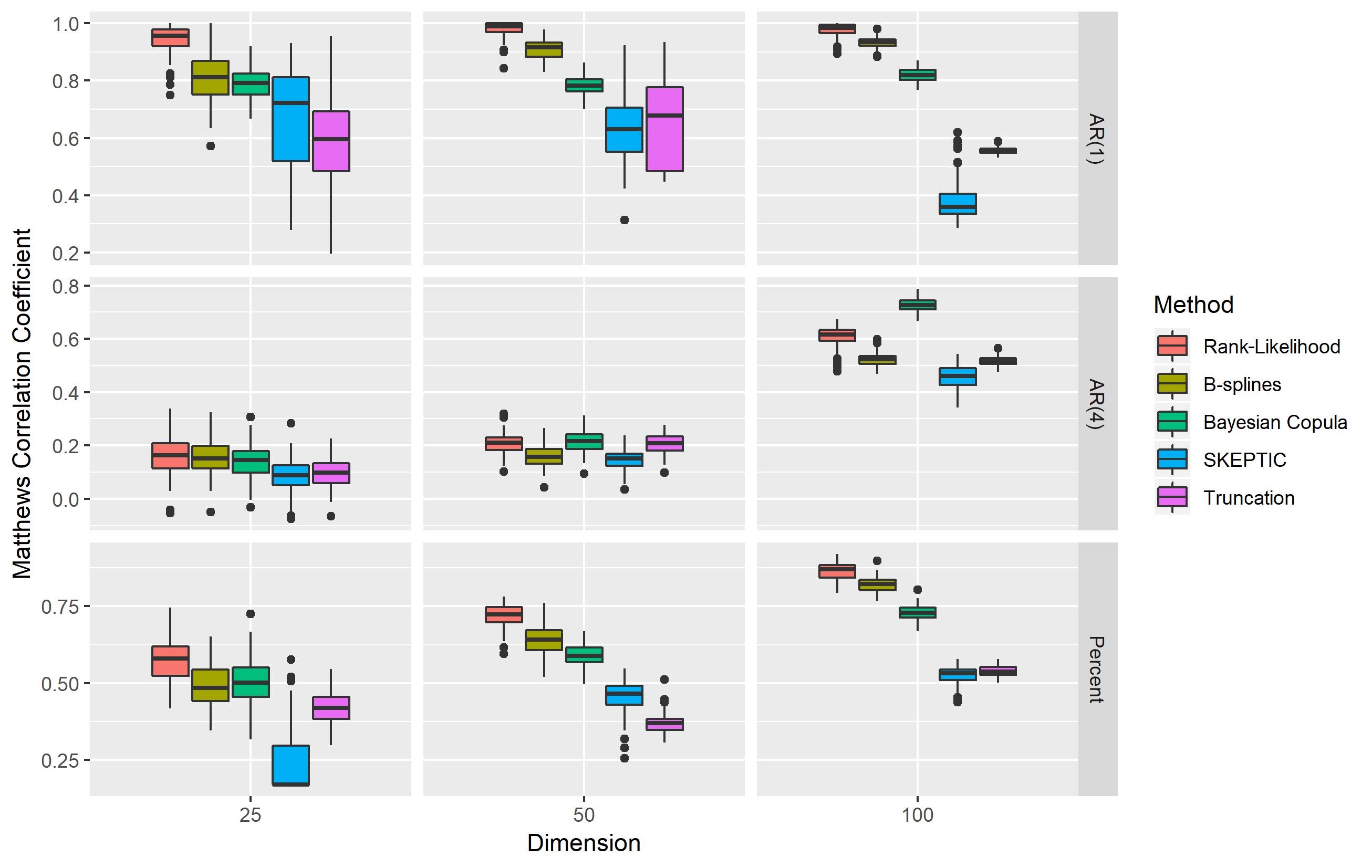}
    \caption{Boxplots of the Matthews correlation coefficient results for each of the methods for different structures of precision matrices. Percent refers to the 10\% model for dimension $p=25$, 5\% model for dimension $p=50$ and 2\% model for dimension $p=100$.}
   \vspace{1ex}
  \end{figure}

    \begin{figure}[!htbp]
    \centering
    \includegraphics[width=.9\linewidth]{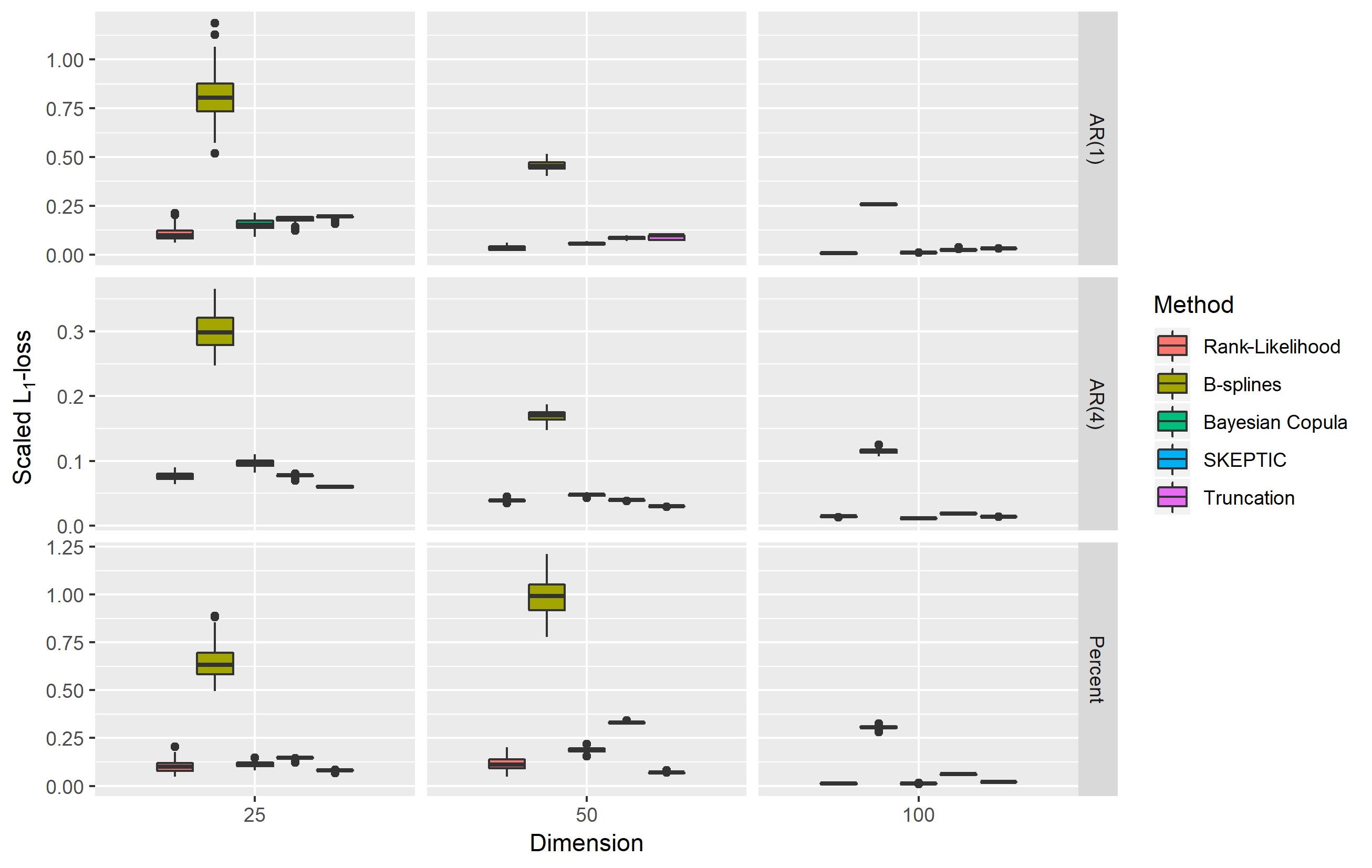}
    \caption{Boxplots of the scaled $L_1$-loss results for each of the methods for different structures of precision matrices. Percent refers to the 10\% model for dimension $p=25$, 5\% model for dimension $p=50$ and 2\% model for dimension $p=100$.}
   \vspace{1ex}
  \end{figure}

\section{Real Data Application}
We demonstrate the methods on a gene expression data set originally referenced in \cite{stranger_population_2007} with Gene
Expression Omnibus database Series Accession
Number GSE6536 19 and its funding is supported in part by the US National Institutes of Health ENDGAME.  Data are collected to measure the gene expression in B-lymphocyte cells from inhabitants in Utah with European ancestry. The interest is on the single nucleotide polymorphisms that are found in the 5' untranslated region of messenger RNA with a minor allele frequency $\geq 0.1$.  Following \cite{bhadra_joint_2013}, of the 47,293 total available
probes, we considered the 100 most variable probes that correspond to different
Illumina TargetID transcripts. The data for these 100 transcripts are available in the {\tt R} package {\tt BDgraph} \citep{mohammadi_bdgraph:_2017, mohammadi_bdgraph:_2019}. The data consist of $n=60$ unrelated individuals and $p=100$ transcripts.  The variables in the data are continuous but do not appear Gaussian.  A Bayesian estimate based on a Gaussian graphical model using a spike-and-slab type prior constructed by \cite{bhadra_joint_2013} detected 55 edges.  

To construct the graph using our method, we convert the original values to be between 0 and 1 using the affine transform $({x - \min(x_i)})/({\max(x_i)-\min(x_i)})$.  We use the identity matrix as the initial matrix for the covariance and inverse covariance matrices for the Rank-Likelihood and B-splines methods.  The Rank-Likelihood method results in 252 edges and the B-splines method results in 99 edges.  Convergence of the Rank-Likelihood method can be obtained in about 28 minutes and for the B-splines method in about 60 minutes for a given $c$ for these data on a laptop computer with an Intel i7 processor and 24 GB of RAM.  Using the same set-up as in the simulation study, the SKEPTIC method results in no edges and the Truncation method results in 363 edges.  The Bayesian Copula method results in 834 edges.  The proposed Rank-Likelihood and B-splines methods result in the sparsest models, with the B-splines method as the most sparse model.  The graphs of the proposed methods are shown in Figure 1.  The graphs of the Bayesian Copula and truncation methods are shown in Figure 2.  Since the SKEPTIC method resulted in no edges, it is not included in the comparison. Plots are made with the {\tt circularGraph} function in {\tt MATLAB}.

\begin{figure}[!htbp]
  \begin{subfigure}[b]{.75\linewidth}
    \centering
  \includegraphics[width=.8\linewidth]{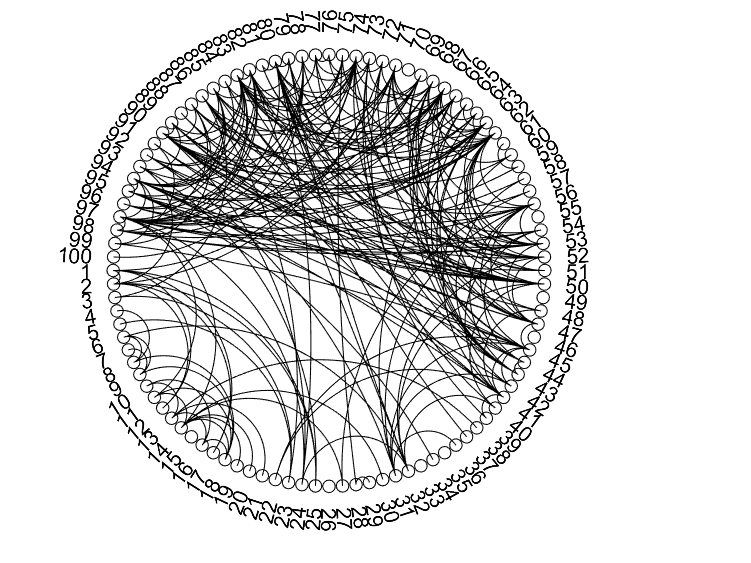}
    \caption{Rank-Likelihood}
   \vspace{1ex}
  \end{subfigure}
  \begin{subfigure}[b]{.75\linewidth}
    \centering
    \includegraphics[width=.8\linewidth]{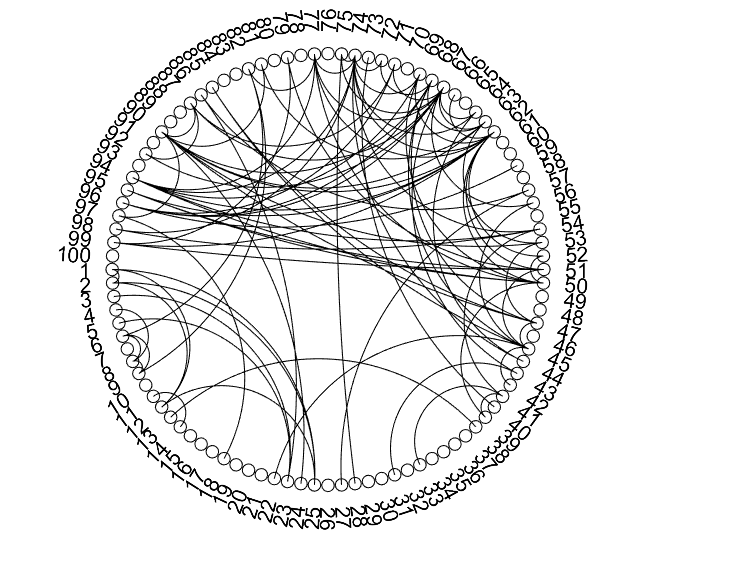}
    \caption{B-splines}
    \vspace{1ex}
  \end{subfigure}
    \caption{Circular graphs illustrating the differences in edges between the methods using the gene expression data set.}
\end{figure}

\begin{figure}[!htbp]
  \begin{subfigure}[b]{.75\linewidth}
    \centering
  \includegraphics[width=.8\linewidth]{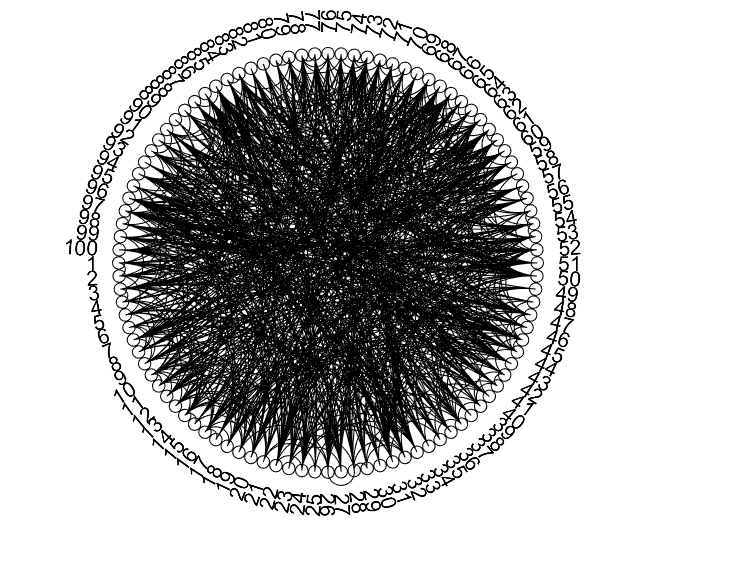}
    \caption{Bayesian Copula}
   \vspace{1ex}
  \end{subfigure}
  \begin{subfigure}[b]{.75\linewidth}
    \centering
    \includegraphics[width=.8\linewidth]{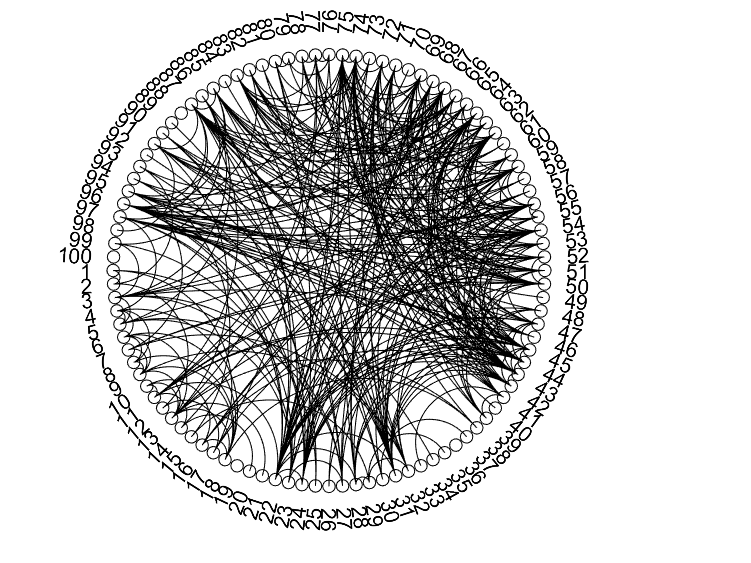}
    \caption{Truncation}
    \vspace{1ex}
  \end{subfigure}
    \caption{Circular graphs illustrating the differences in edges between the methods using the gene expression data set.}
\end{figure}

\bigskip
\begin{center}
{\large\bf SUPPLEMENTAL MATERIALS}
\end{center}

\begin{description}
\item[GitHub Repository:]
The code for the methods described in this paper can be found in the following GitHub repository: \url{https://github.com/jnj2102/RankLikelihood}.

\end{description}

\section*{References}

\bibliography{Zotero.bib}

\end{document}